\documentclass{XrU2005}
\usepackage{graphicx}
\usepackage{natbib}

\title{X-ray constraints on the dark matter profile of a very relaxed
  cluster of galaxies}
\author[1]{L. Zappacosta}
\author[1]{D. A. Buote}
\author[1]{F. Gastaldello}
\author[1]{P. J. Humphrey}
\author[1]{J. Bullock}
\affil{Department of Physics and Astronomy, University of California,
  Irvine}
\author[2,3]{F. Brighenti}
\affil[2]{UCO/Lick Observatory, Board of Studies in Astronomy and
  Astrophysics, University of California, Santa Cruz}
\author[2]{W. Mathews}
\affil[3]{Dipartimento di Astronomia, Universit\`a di Bologna,
  Bologna, Italy}

\begin{document}

\keywords{ X-rays: galaxies: clusters; dark matter;  galaxies: clusters}

\maketitle

\begin{abstract}
\vspace{-0.4cm}
We have analyzed an XMM-Newton observation of the cluster Abell~2589. 
Apart from a low-level asymmetry in the central region, the cluster 
appears very relaxed and does not show presence of a
central AGN. We derived constraints for the radial temperature,
density and, assuming hydrostatic equilibrium, mass profiles. We find
that the best fit to the dark matter profile is given by the
Sersic-like profile proposed by \citet{navarro04}. The NFW model does
not provide a good fit. We also tested whether the central stellar
component could affect the profile through the adiabatic contraction model
but were unable to distinguish it from a simple dark matter + stars
modeling.
\end{abstract}
\vspace{-0.4cm}
\section{Introduction}
\vspace{-0.4cm}
Clusters of galaxies, being the largest bound and dark
matter dominated objects in the universe, are an optimal place to test
the predictions of cosmological simulations regarding the mass profile
of their dark halos. In this regard their X-ray emission can be
successfully used to constrain the mass profile as long as the
emitting plasma is in hydrostatic equilibrium. For this reason, to
compare with theoretical predictions, we need to study very relaxed
systems that do not show any sign of disturbance in their
morphology. Clusters like these are very rare since they often show
signs of interactions with other objects and, especially the more
relaxed ones, almost always show a central radio galaxy whose
influence on the hot plasma can easily invalidate the assumption of
hydrostatic equilibrium. Here we show the results of an XMM-Newton
analysis of Abell~2589 (z=0.0414), a very relaxed cluster with no
presence of central radio emission.
\vspace{-0.4cm}
\section{Data reduction}
\vspace{-0.4cm}
The data reduction was performed using SAS 6.0. We
excluded the point sources by first looking at the PPS source list and
then through a visual inspection. The $50\,\rm{ksec}$ observation we have
analysed was affected by frequent periods of strong flaring. 
Having screened the data, based on light curves from a ``source-free''
region in different energy bands, the final exposure times were 
17~ksec and 13~ksec
respectively for the MOS and PN detectors. We modeled the background
by performing a simultaneous fit of the spectra of the outermost 4 annuli
we have chosen for the spectral analysis.
\vspace{-0.4cm}
\section{Spatial analysis}
\begin{figure}[t]
\includegraphics[width=0.45\textwidth, angle=0]{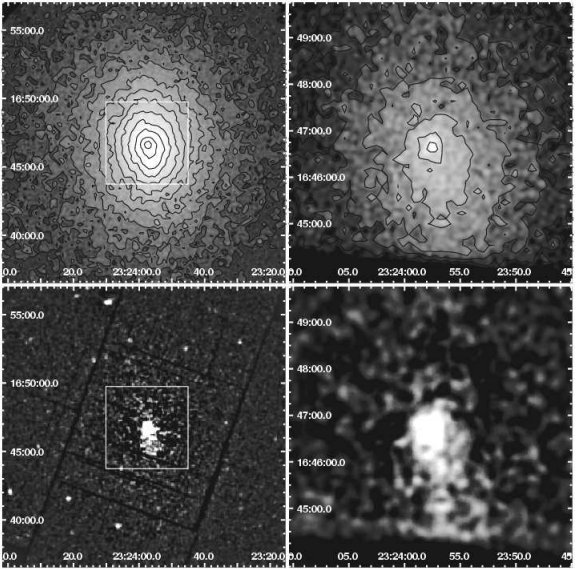}
\caption{Upper panels: XMM-MOS and Chandra images. Lower panels:
  XMM-MOS1 and Chandra unsharp mask images. The XMM images report
  the Chandra field 
of view.\label{imgs}}
\end{figure}

\vspace{-0.4cm}
In Fig.~\ref{imgs} we show the XMM-MOS and Chandra X-ray
images\footnote{The Chandra images are from a 14~ksec observation 
previously analized by \citet{buote}.} of the cluster and
their unsharp mask images obtained by differencing images smoothed by
gaussian kernels of $5^{\prime\prime}$ and $40^{\prime\prime}$. 
The images show very regular isophotes
with ellipticities of $\sim0.3$. The only disturbance in the morphology is a
southward centroid offset very well shown in the unsharped mask
images. This offset region has an emission only 30\% higher than the mean
cluster emission at $\sim60\,\rm{kpc}$ from the center corresponding
to $\sim 15\%$ variation in the gas density. We also produced an
hardness ratio map and could not find any significant non radial
variation in temperature. The cluster has a central dominant bright
galaxy centered at the X-ray peak. \citet[][]{beers} measured its
relative velocity finding that is unusually high for a dominant
galaxy. The distribution of galaxies shows a preferential north-south
alignment (2.5 degrees to the south there is Abell~2593) and a big
subclump to the north (in the opposite side of the X-ray offset)
off-centered by $3^{\prime}$ from the X-ray peak.  The well relaxed gas phase
appearance and the particular galaxy distribution may be revealing a
mild process of accretion through the large-scale structure
\citep{plionis} that does not greatly disturb the gas properties.
\vspace{-0.4cm}
\section{Spectral analysis}
\vspace{-0.4cm}
We extracted spectra from 7 concentric annuli centered on the X-ray
peak and obtained gas density and temperature\footnote{We fitted APEC
 models modified by the Galactic absorption using XSPEC.} profiles.  We have
analyzed only the projected quantities that with the quality of our
data give us the best constraints.  
The best fit to the projected gas density is obtained using a 
cusped $\beta$ model with
core radius $r_c=110\pm12\,\rm{kpc}$, cusp slope $\alpha=0.3\pm0.1$
and $\beta=0.57\pm0.01$. A single $\beta$ model does not fit the inner
two data points. The temperature profile of Abell~2589 is 
almost isothermal as already shown by the Chandra analysis of
this object by \citet{buote}. The important deviations from
isothermality are in the inner and outer data points that have lower
temperatures. The resulting profile has been parametrized using two
power-laws joined smoothly by exponential cut-offs.
\begin{figure}[!t]
\includegraphics[width=0.46\textwidth]{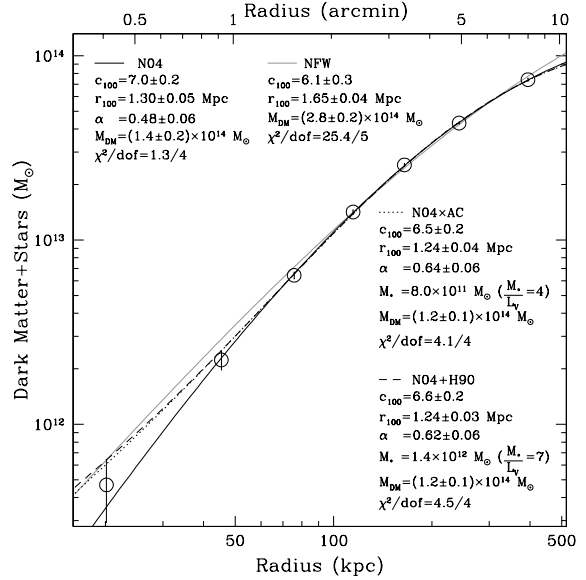}
\caption{{\em Dark matter+stars} profile. The models discussed in
  Sect.~\ref{dm} are reported. 
The virial quantities refer to a 
 halo whose mean density is $100\,\rho_c$.\label{dm+stars}}
\end{figure}
\vspace{-0.4cm}
\section{Dark matter profile}\label{dm}
\vspace{-0.4cm}
Given the parametrized quantities we can calculate the {\em total
gravitating mass} profile (assuming hydrostatic equilibrium) and infer
constraints on the dark matter profile.  The {\em dark matter+stars} profile
($\mathit{total\ mass - gas\ mass}$) and the fitted models are shown in 
Fig.~\ref{dm+stars}. 
The NFW profile \citep[solid grey line;][]{navarro} 
is a good fit except for $\rm{r}<80\,\rm{kpc}$. 
The updated Sersic-like CDM profile proposed by
\citet{navarro04} (hereafter N04) is able to provide a good fit 
to the entire {\em dark matter+stars} profile.
We tried to
assess the level of importance of the stellar component due to the
central bright galaxy, modeled with an Hernquist profile
\citep[][hereafter H90]{hernquist}, using parameters from \citet{malumuth}. 
We also tested the influence of baryonic condensation into stars by using
the adiabatic contraction model (AC) of \citet{gnedin}. If we let
the total mass in stars $\rm{M_*}$ be free to vary, the data do not require
any stellar component. If we fix $\rm{M_*/L_v}$ we can still obtain a
reasonable fit allowing for $\rm{M_*/L_v}=7$ in case of a N04+H90 profile
(dashed black line) and $\rm{M_*/L_v}=5$ in case of a N04 with adiabatic
contraction ($\rm{N04\times AC}$; dotted black line). In general we are not able to
discriminate between models with and without adiabatic contraction.
\vspace{-0.4cm}
\section*{Acknowledgments}
\vspace{-0.4cm}
We thank O. Gnedin for providing us the code for the adiabatic contraction.
%
%
%

%
\end{document}